\documentclass[10pt,letterpaper]{article}
\usepackage{cite}
\usepackage{graphicx}
\usepackage{fullpage}

\begin{document}
\title{High temperature epsilon-near-zero and epsilon-near-pole metamaterial emitters for thermophotovoltaics }
\author{Sean Molesky, Christopher J. Dewalt, and Zubin Jacob* \\University of Alberta, Department of Electrical and Computer Engineering \\ T6G 2R3 Edmonton, Alberta, Canada \\ \small{\underline{*zjacob@ualberta.ca}}}
\maketitle
\begin{abstract}We propose a method for engineering thermally excited far field electromagnetic radiation using epsilon-near-zero metamaterials and introduce a new class of artificial media: epsilon-near-pole metamaterials. We also introduce the concept of high temperature plasmonics as conventional metamaterial building blocks have relatively poor thermal stability. Using our approach, the angular nature, spectral position, and width of the thermal emission and optical absorption can be finely tuned for a variety of applications. In particular, we show that these metamaterial emitters near 1500 K can be used as part of thermophotovoltaic devices to surpass the full concentration Shockley-Queisser limit of 41\%. Our work paves the way for high temperature thermal engineering applications of metamaterials.
\end{abstract}
\setlength\intextsep{-100pt}
\setlength\belowcaptionskip{-10pt}


\section{Introduction}
High temperature energy conversion processes, such as combustion, are accompanied by thermal losses which often outstrip usable produced power \cite{Taylor2008}. As an example, for average oil and coal based energy conversion, thermal losses account for roughly 50\% to 60\% of the total produced power \cite{Buskies1996,Rosen2003}. For small internal combustion engines this number may climb as high as 75\% \cite{Chaisson2008}. These losses can be substantially mitigated by utilizing the waste thermal energy as a heat resource. An ideal set up would convert waste thermal energy directly into usable power.

Thermal losses are also significant for photovoltaic cells. Due to the large spectral range that must be captured, over 60\% of incident solar power is lost as thermal energy \cite{Datas2010a}. Regardless of the material bandgap's spectral position, radiation above the bandgap loses a portion of its energy to thermalization within the cell. Radiation with energy below the bandgap is essentially unused. These effects are intrinsic and place an upper limit on energy conversion for a single junction semiconductor cell \cite{Shockley1961}. While proposals of multi-junction cells offer a potential workaround, they also require a shift to expensive materials and structures. 

An alternative solution exists in the application of the thermophotovoltaic (TPV) method \cite{Wedlockt1963}. In this approach, any combination of conductive, convective or radiated waste energy is concentrated to heat a structure with a spectral emission tailored to match the bandgap of a specific photovoltaic cell. Once the structure is heated, the source energy is converted to electromagnetic radiation which can be transformed into electric power with high efficiency. The method of maintaining the heated source is flexible giving the TPV approach applicability beyond that of large scale solar or solar thermal approaches. For example, TPV methods can be used to create compact devices for cogeneration of heat and electricity.

TPV devices can easily be integrated as efficiency increasing components of larger systems, or function as primary solid state energy converters \cite{JXcrystals}. Unfortunately, the operational temperatures needed for sufficient power generation and efficient energy conversion \cite{Gee2002} as well as our limited ability to tune the thermal radiation spectrum have made previous TPV approaches largely impractical. While excellent progress has been made towards designing emitters for far field TPVs through surface structuring \cite{Mason2010,Wu2012,Mason2011,Maksimovic2008,Liu2011,Bermel2011,Han2010,Chen2007}, planar Fabry-Perot based structures \cite{Nefzaoui2012,Chang2011,Rephaeli2009} and photonic crystals \cite{Schuler2009,Li2002,Celanovic2004}, a dominant, broadly applicable set of tools for controlling thermally induced radiation has yet to emerge. 

In this paper, we introduce a class of thermal effects in metamaterials for thermophotovoltaic applications. Our approach rests on engineering the poles and zeros of the dielectric constant which allow for an array of unique optical responses. We show that epsilon-near-zero (ENZ) metamaterials can behave as spectrally tunable, narrowband ultra-thin thermal emitters and introduce a new class of artificial media, epsilon-near-pole (ENP) metamaterials. This class of ENP media are ideally suited as thermal emitters in a thermophotovoltaic system where the main requirements are 1. omnidirectional thermal emission, 2. narrowband and high emissivity, and 3. polarization insensitivity. We also address one of the major limitations of conventional metamaterials for thermal applications: high temperature operation; by switching to plasmonic materials with high melting points. Finally, we analyze the performance of these emitters within a practical thermophotovoltaic device and show that the energy conversion efficiency can exceed the Shockley Queisser limit of single junction cells. Note that the ENZ and ENP thermal emitters introduced in this paper can function as ideal narrowband ultra-thin absorbers as well. 

\section{Epsilon-near-zero (ENZ) and epsilon-near-pole (ENP) narrowband absorbers and emitters}
\subsection{Tailored emission and optical absorption}
Making use of Kirchhoff's law of thermal radiation for a body in thermodynamic equilibrium \cite{Greffet1998}, the engineering of thermal emission can be formulated in terms of optical absorptivity:
\begin{equation}
\zeta\left(\lambda,\theta,\phi\right)=\alpha\left(\lambda,\theta,\phi\right),
\label{Kirchhoff}
\end{equation}
with $\alpha\left(\lambda,\theta,\phi\right)$ denoting the structure's absorptivity as function of wavelength, azimuthal angle, and polar angle, and $\zeta$ the structure's emissivity.  It follows directly that spectrally narrow regions of high optical absorption also create spectrally narrow regions of high thermal emission. Consequently, the use of optical resonances provides a natural starting point for designing thin structures to control thermally excited electromagnetic radiation.  We introduce the concept of thermal engineering using ENZ and ENP resonances and show how the fundamentally distinct natures of these two bulk material resonances can create a range of thermally induced effects. The proposed approach can be used for a variety of applications and in particular for TPVs, where the main constraints on the emitter are narrowband and omnidirectional emissivity \cite{Datas2010}. 
\subsection{ENZ and ENP absorption}
\underline{ENZ}: Lossless or near lossless epsilon-near-zero resonances have recently been a topic of broad research interest, and have been shown as a plausible mechanism for creating high performance optical devices ranging from nonlinear optical switches to tailored radiation phase patterns \cite{Wurtz2011,Alu2007}. We show here that ENZ resonances have important applications in general control of thermally induced radiation (explored in Sec.III). However, the traditional $Re\left(\epsilon\right)\rightarrow 0$ and $ Im\left(\epsilon\right)\rightarrow 0$ ENZ regime is not suited to the requirements of a TPV emitter. 

P-polarized radiation incident on an ENZ slab shows increased absorption (non-normal incidence). This resonance arises due to the presence of a field enhancing mechanism that relies on the  displacement field boundary condition: $\epsilon_{1}E_{1\perp}=\epsilon_{2}E_{2\perp},$ where the $\perp$ denotes the direction perpendicular to the slab, and either medium can be assumed to have ENZ behavior (if $\epsilon_{2} \rightarrow 0$ then $E_{2\perp} \rightarrow \infty$).  Kirchoff's laws immediately reveals that this ENZ slab with enhanced absorption should show a high emissivity.  However, s-polarized light which does not have a component of the field perpendicular to the slab does not show this field enhancement or the ENZ resonance.   By this constraint, no s-polarized light can be thermally excited, and p-polarized radiation cannot be efficiently emitted at low polar angles. Since the emission of an ideal blackbody shows no angular or polarization preference, the maximal averaged emitted spectral power in an ENZ region is less than half of what can be achieved theoretically.


Nevertheless, polarization averaged emissivity near that of a blackbody can still be attained if the $Im\left(\epsilon\right) \rightarrow 0$ condition is relaxed. In moving away from true ENZ behavior by the addition of extra loss, two separate, but connected, absorption improving effects occur. First, in the $Re\left(\epsilon\right)\rightarrow 0$ region, the high impedance mismatch between an ENZ material and free space is greatly reduced as the added loss acts to decrease the impedance of the material. Since this also dictates a general relaxation of ENZ resonance characteristics, polarization sensitivity is greatly diminished. Second, at wavelengths shorter than the $Re\left(\epsilon\right)\rightarrow 0$ crossing where material impedance is similar to that of free space ($Re\left(\epsilon\right)\approx 1$), the addition of material losses begins to allow for significant absorptivity even if the material film is thin. Both effects push this pseudo ENZ resonance towards near-omnidirectional and high absorptivity for both electromagnetic polarizations.  At higher losses, they combine to create a single highly absorptive spectral region (Fig. 1(A)). 

Yet, improving absorptivity in this manner comes at the cost of an increased spectral width. Due to the natural dispersion limitations of a region where $Re\left(\epsilon\right)\rightarrow 0$, the spectral width over which the additional loss achieves impedance matching is comparatively broad. As a direct result, high emissivity occurs over a much wider range than that ideal for high efficiency TPVs. The onset of this behavior is depicted in Fig.1(A), where the emissivity of a 100 nm thick film moving away from the true ENZ characteristics, following a Drude model, on a perfectly reflecting backing is calculated using (\ref{Kirchhoff}) and the Poynting vector relation $ \vec{S} = \frac{1}{2} \vec{E} \times \vec{H}^{*}$. Again, as will be discussed in Sec.V, while broader emissivity may be useful for certain TPV applications, it does not match the ideal narrowband criterion. In light of these results, we introduce the concept of ENP resonances for achieving the thermal emission characteristics necessary for high efficiency TPVs. Note that both the ENZ and ENP resonances can be engineered using nanostructured metamaterials.
\\
\begin{figure}
\centering
\includegraphics{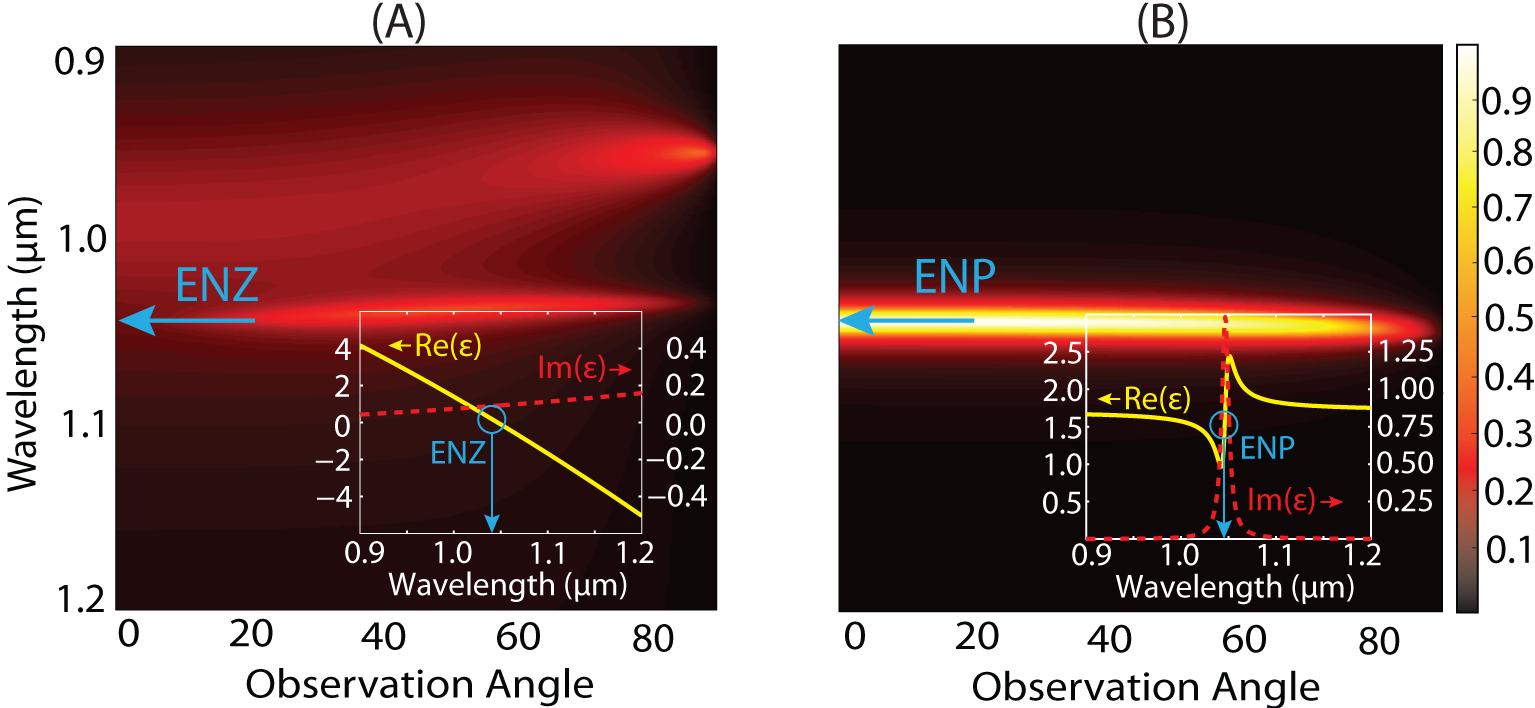}
\caption{: (A) Polarization averaged emissivity of a 100 nm thick film of ENZ material, following a Drude model, on a perfectly reflecting backing. The spectrally narrow peak is the ENZ resonance. The broad blur at shorter wavelengths is the onset of impedance matching in the $Re\left(\epsilon\right)\approx 1$ range.  As loss is increased, these regions blend together and the magnitude of the emissivity tends towards unity in a broad spectral range. The inset shows the relative dielectric constant used. (B) Polarization averaged emissivity of a 100 nm thick material film in the ENP regime, based on a Lorentz model, on a perfectly reflecting backing. The component loss considered is identical to that used in Fig. (A): $\gamma_{A}=\gamma_{B}$. Again, the inset shows the relative dielectric constants used to calculate the polarization averaged emissivity. The spectrally sharp behavior shown here makes ENP type resonances a promising candidate for TPV applications.}
\end{figure}

\underline{ENP}: The primary advantage of operating at an ENP resonance is the extremely dispersive nature of these regions. This characteristic allows for tight spectral control even with moderate material losses. Yet, beyond this most important feature, several other benefits for TPV type emitters and absorbers exist. At an ideal pole of the dielectric constant we have, $Re\left(\epsilon\right) \rightarrow \left(\pm\right)\infty$. The addition of losses regularizes the singularity, reduces impedance mismatch and leads to enhanced absorption (high emissivity) in a narrow spectral region (Fig 1.(B)). The ENP resonance associated with such a pole shows no polarization sensitivity and achieves omnidirectional high emissivity in isotropic media. 


We now compare the critical role of losses in the ENZ and ENP resonance.  In contrast to the fixed $Re\left(\epsilon\right) \rightarrow 0$ condition of an ENZ resonance, which forces any impedance matching to be accomplished only through the addition of loss, the behavior of $Re\left(\epsilon\right)$ itself in an ENP region is highly loss dependent. When loss is added, the real part of the dielectric function can no longer achieve large absolute values, opening the possibility of impedance matching even with relatively lower losses as compared to the ENZ resonance. This leads to a narrower spectral window of high emission and absorption than the previously mentioned ENZ approach. A comparative visualization of the characteristic spectral emissivity of a 100 nm thick material film in the ENP regime with moderate losses, based on a Lorentz model, on a perfectly reflecting backing is shown in Fig.1(B). Noting that ENP type resonances match all the requirements of a high efficiency TPV emitter, we now focus on practical realizations of ENZ and ENP materials. 

\section{Practical ENZ and ENP with metamaterials}
\subsection{Natural ENZ and ENP resonances}
In natural materials, ENZ regions occur at bulk plasmon as well as longitudinal optical phonon resonances, while ENP characteristics are related to transverse optical phonons. Yet, despite the ubiquity of these features in optical responses, few materials exhibit ENZ or ENP characteristics in the 0.5 eV to 1.0 eV range, crucial for TPV devices. The bulk plasmon energy, proportional to $\omega_{p}\propto \left(N/m_{e}\right)^{1/2}$, is generally pushed to much higher energies due to the small effective electron mass, $m_{e}$, and the high electron concentration, $N\approx 10^{22}cm^{-3}$, of typical metals \cite{Ashcroft1968}. The energy of material phonon resonances, proportional to $ \omega_{LO}\propto\omega_{TO}\propto \left(1/M\right)^{1/2}$, occurs at significantly lower energies due to the relatively large reduced ionic mass, M \cite{Ashcroft1968}. The prospect of natural ENP or ENZ infrared emitters is limited to a small collection of highly lossy materials such as osmium or molybdenum which are not capable of creating the spectrally narrow emission required for high efficiencies.

To engineer ENP and ENZ responses we consider two approaches based on metamaterial crystals. The following subsections show how ENP and ENZ features beyond those found naturally can be created and tuned through the use of either planar material stacks or embedded nanowires. Schematics of the multilayer and nanowire structures are depicted in Fig.2.
\begin{figure}
\centering
\includegraphics{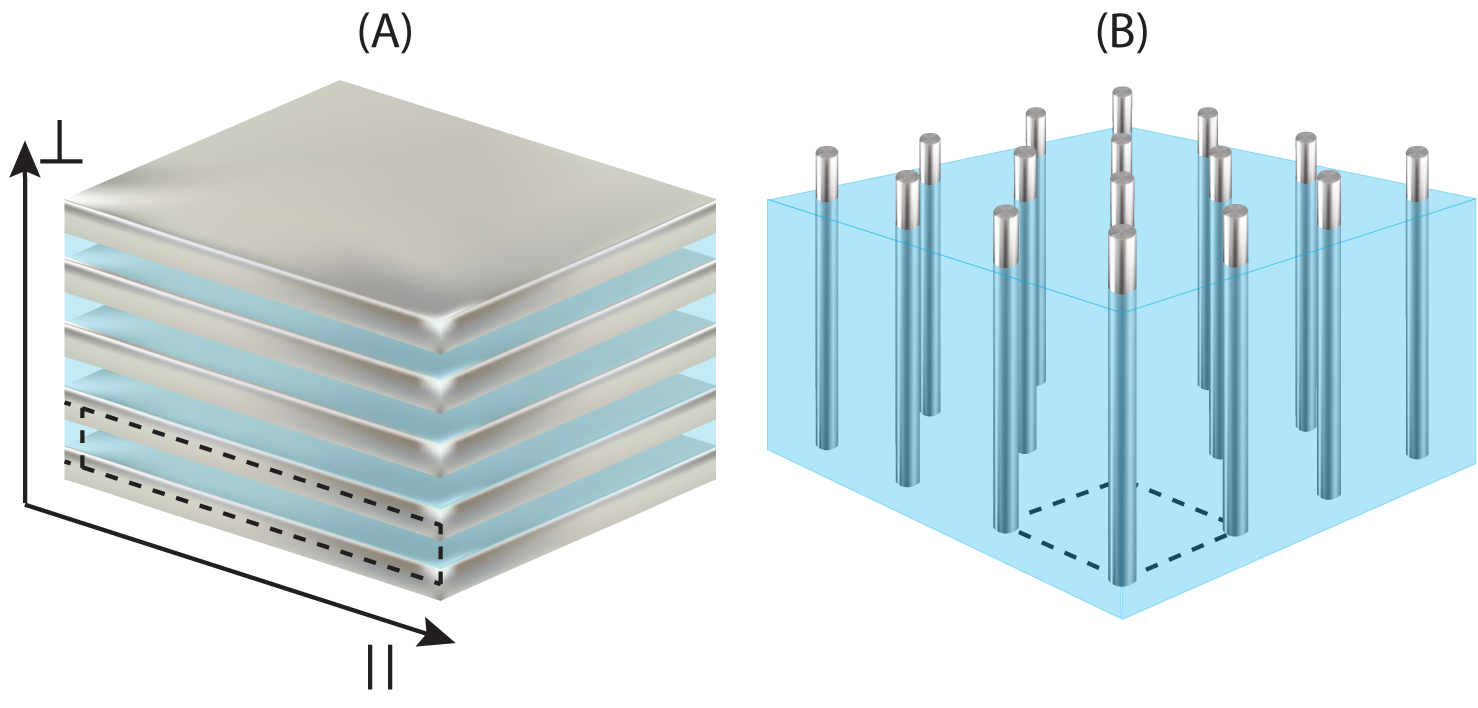}
\caption{: (A) Schematic of a multilayer metamaterial created by interlacing layers of optical metal and dielectric. (B) Schematic of a nanowire metamaterial created by embedding metallic nanowires in a host dielectric matrix. Both structures can be created with current fabrication techniques \cite{Yao2008,Wurtz2011}.}
\end{figure}
\subsection{One-dimensional multilayer structure}
The simplest structure for creating the ENZ/ENP metamaterials consists of alternating layers of metal and dielectric forming a multilayer structure (Fig. 2(A)).  The effective medium parameters are given by
\begin{equation} 
\epsilon_{||}=\epsilon_{M}\rho+\epsilon_{D}\left(1-\rho\right)
\quad
\epsilon_{\perp}=\frac{\epsilon_{M}\epsilon_{D}}{\epsilon_{M}\left(1-\rho\right)+\epsilon_{D}\rho},
\label{planar}
\end{equation}
where the M subscript is used to denote the metal, the D subscript the dielectric and $\rho$ the relative fill factor of metal in the unit cell. The parallel ($||$) and perpendicular ($\perp$) subscripts show the direction convention that will be used in the remainder of the text.

The physical characteristics of these effective parameters can be understood in terms of the restrictions placed on the material's electrons. Since the motion of electrons is nearly free within each metal plane (Fig.2(A)) the parallel effective medium parameter follows the frequency dispersion relation of an effective metal. In turn, this creates ENZ behaviour at the effective plasmon resonance of the parallel direction. Conversely, in the perpendicular direction nearly free electrons are confined to the thickness of their particular metal plane. On account of this confinement, the frequency dispersion of the perpendicular permittivity mirrors that of an effective Lorentz model dielectric with an accompanying effective ENP resonance. The performance of these effective medium parameters, for calculating emissivity, is compared with the exact theoretical transfer matrix technique for a model tantalum/titanium dioxide system in Fig.3. Note that throughout the paper, we verify the predictions of effective medium theory (EMT) using numerical simulations taking into account the finite size of the unit cell, absorption in the constituent materials, dispersion as well as effects due to the substrate.  
\begin{figure}
\centering
\includegraphics{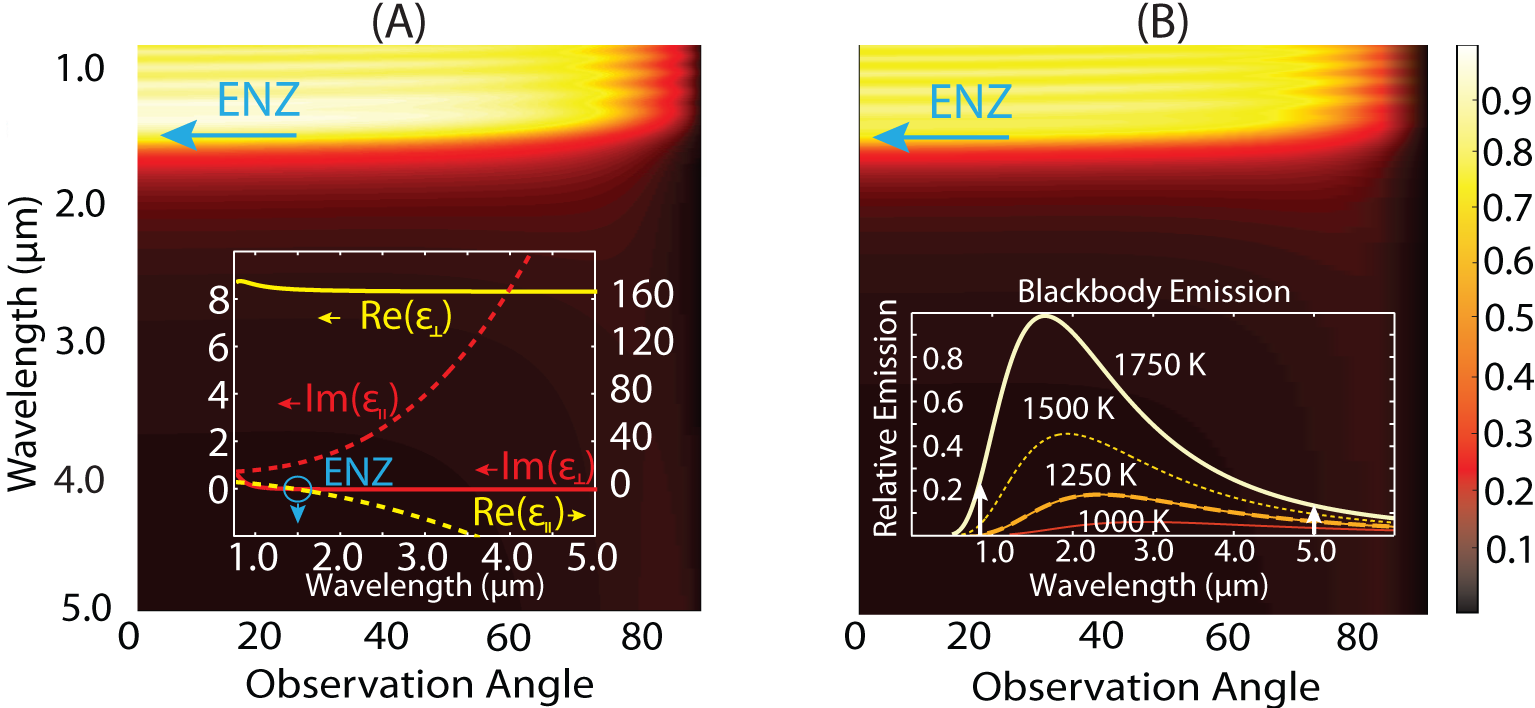}
\caption{Multilayer ENZ emitter: (A) Effective medium theory calculation of the emissivity of a planar multilayer structure. The metamaterial is composed of twenty unit cells of 5 nm thick tantalum (modeled by a Drude relation) and 45 nm of titanium dioxide ($\epsilon = 7.5$) on optically thick tantalum. Both materials can be deposited by atomic layer deposition \cite{Lim2003}. The inset shows the effective medium parameters as functions of wavelength. The ENP resonance is located outside of the plotted area and has little effect due to the spectral power distribution of a blackbody. (B) Transfer matrix calculation of the multilayer structure which shows excellent agreement with EMT. The inset shows the relative emission strength of an ideal blackbody as function of wavelength. The arrows denote the cutoffs of the emissivity plots.}
\end{figure}

Taking the dispersion characteristics into account, the limitations of the planar structure for thermal applications become apparent. As ENP behavior occurs along the optic axis, its interaction with light is limited to the p-polarization. For realistic materials, this limited domain of interaction leads only to a small lobe of enhanced emissivity at high polar angles. In the parallel direction, the ENZ resonance displays the same characteristics as an isotropic ENZ material. Since the parallel optical response parameter interacts strongly with both polarizations, ENZ behavior dominates the overall emissivity features of the multilayer structure. Correspondingly, the same broad emissivity that earlier caused us to turn toward ENP behavior is also a general feature of the multilayer structure. A visualization of the emissivity of a realizable multilayer system using model tantalum and titanium dioxide is shown in Fig.3. Possible application of this structure as an emitter for TPVs will be explored in Sec.V.

\subsection{Two-dimensional nanowire structure}

Seeking spectral emissivity control beyond what can be accomplished using the multilayer structure, we switch to the nanowire metamaterial system (Fig. 2(B)) \cite{Pollard2009, Yao2008}. In contrast to the multilayer structure, embedded nanowires allow nearly free electron propagation along the perpendicular direction (the optic axis), but not in the parallel plane. As the confinement of electrons is flipped from the multilayer structure, so too is the functional frequency dispersion. The perpendicular direction now follows the frequency dispersion characteristics of a metal, while the parallel mimics that of an effective Lorentz model dielectric. Employing the same generalized Maxwell-Garnett approach used for the planar metamaterial, the effective medium parameters for metallic nanowires with a square lattice embedded in a dielectric matrix \cite{Elser2006} are defined as:
\begin{equation}
\epsilon_{||}=\epsilon_{D}\left[\frac{\epsilon_{M}\left(1+\rho\right)+\epsilon_{D}\left( 1-\rho\right)}{\epsilon_{M}\left(1-\rho\right)+\epsilon_{D}\left(1+\rho\right)}\right]
\quad
\epsilon_{\perp}=\rho \epsilon_{M}+\left(1-\rho\right)\epsilon_{D},
\label{nanowire}
\end{equation}
where we have used the same convention earlier introduced for the multilayer structure.
Following the general dispersion relations of the effective medium parameters, shown in the inset of Fig. 4(B), the ENP and ENZ conditions are now properly aligned for TPV applications. The ENP resonance which results in high narrowband emissivity is located along the parallel direction where it interacts with both s- and p-polarized light. As such, the general emissivity pattern of the nanowire structure is very similar to that of an isotropic ENP material, and nearly ideal for creating a TPV emitter. This is shown in Fig.4, where titanium nitride nanowires embedded in silicon have been used to create a spectrally narrow emissivity spike at a wavelength just below the bandgap of a gallium antimonide (GaSb) photovoltaic cell. The spectral location of GaSb's bandgap, 1.75 $ \mu$m, makes it well suited for use in a TPV device at realistic operating temperatures (inset Fig. 3(B)).  
\begin{figure}[ht!]
\centering
\includegraphics{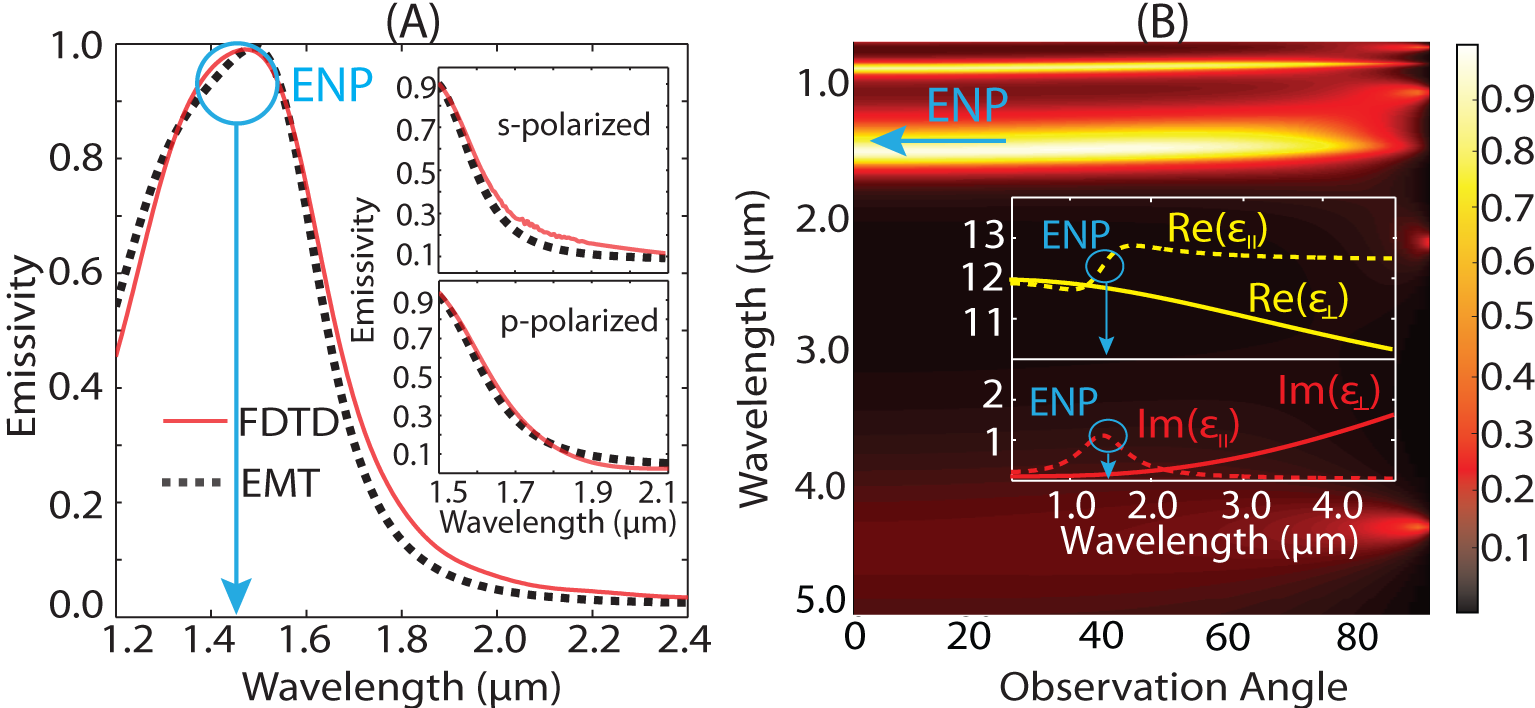}
\caption{Nanowire ENP emitter: (A) Comparison of the polarization averaged emissivity of a 280 nm thick metamaterial emitter making use of a host matrix of silicon (assumed to be a constant dielectric) and 20 nm diameter titanium nitride nanowires in a 120 nm square unit cell on an optically thick tantalum backing. The two curves compare emissivity as calculated by effective medium theory and finite difference time domain simulation (Lumerical) at normal incidence. The insets show the same comparison for s- and p-polarized emissivity over a compressed wavelength range of 1.5 to 2.1  $ \mu$m at a polar angle of 50 degrees. Note the excellent agreement between EMT and the full numerical simulation. (B) Polarization averaged emissivity of the nanowire system described in (A) calculated using EMT. Emission peaks occurring below the designed emission are known to be part of the Bragg scattering regime \cite{D'Aguanno2012}. These peaks have little effect in application due to low emitted power at wavelengths shorter than 800 nm for bodies cooler than 3000 K. The inset shows the effective medium parameters as functions of wavelength. The spectrally narrow, omnidirectional nature of the ENP emissivity peak is nearly ideal for use as an emitter in a TPV device.}
\end{figure}
\subsection{Metamaterial thermal antenna}
We now show how the ENZ region of the nanowire array can lead to highly directional thermal emission. Primarily, this occurs as the optical dispersion relation, relating the energy and momentum of an electromagnetic wave in a medium, is altered from the isotropic form of $k_{\perp}^{2}+k_{||}^{2}=\epsilon\omega^{2}/c^{2}$ (where we have used the previously mentioned direction convention, \emph{k} denotes the spatial frequency of the wave and $\omega$ its frequency) to the anisotropic form of:
\begin{equation}
\frac{k_{\perp}^{2}}{\epsilon_{||}}+\frac{k_{||}^{2}}{\epsilon_{\perp}}=\frac{\omega^{2}}{c^{2}}.
\label{aniso}
\end{equation}
 If $\epsilon_{\perp}\ne \epsilon_{||}$, $Re\left(\epsilon_{\perp}\right)>0$ and $Re\left(\epsilon_{||}\right)>0$ the isofrequency surface (generated by fixing $\omega$ and allowing $k_{\perp}$ and $k_{||}$ to vary) forms an ellipsoid. If either $Re\left(\epsilon_{\perp}\right)<0$ or $ Re\left(\epsilon_{||}\right)<0$, with the real part of the other dielectric component $>0$, the dispersion relation forms a hyperboloid.

As the perpendicular direction nears low loss ENZ behavior, the isofrequency surface of the anisotropic crystal, Eq.(\ref{aniso}), with $Re\left(\epsilon_{||}\right)>0$ becomes very narrow (Fig.5). Accordingly, very small changes in incident angle (parallel wave momentum) result in extreme variation of the material impedance. The electromagnetic wave begins to probe the perpendicular ENZ component at non-normal incidence and material impedance becomes very large leading to high reflections \cite{Alekseyev2010}. This allows a small angular region (for p-polarized light) over which the wave can penetrate the medium and be absorbed (Fig. 5). Thus the metamaterial can show high emissivity only over this narrow angular region at the ENZ wavelength. Concurrently, the closer $Re\left(\epsilon_{\perp}\right)$ is to 0, the narrower this region becomes, eventually forming a needle-like angular peak. Since spatial coherence is directly connected to the angular spread of an electromagnetic wave, the heating of an anisotropic crystal in the region of a perpendicular ENZ resonance presents a grating-free method of thermally exciting coherent radiation. The angular and spectral location of this antenna-like behavior is tunable based on the fraction of metal, $\rho$, filling the underlying unit cell. It is worth mentioning that in the extremely low loss ENZ case a second, purely longitudinal, electric field solution is possible \cite{Agarwal1974,Pollard2009}.  Regardless, the general behavior of the low loss ENZ results shown in Fig.5(C) is still possible with the inclusion of this second solution. For TPV applications, omnidirectional emission achievable at the ENP wavelength is preferable and a more detailed treatment of the ENZ thermal antenna will be presented elsewhere. 

\begin{figure}[ht!]
\centering
\includegraphics{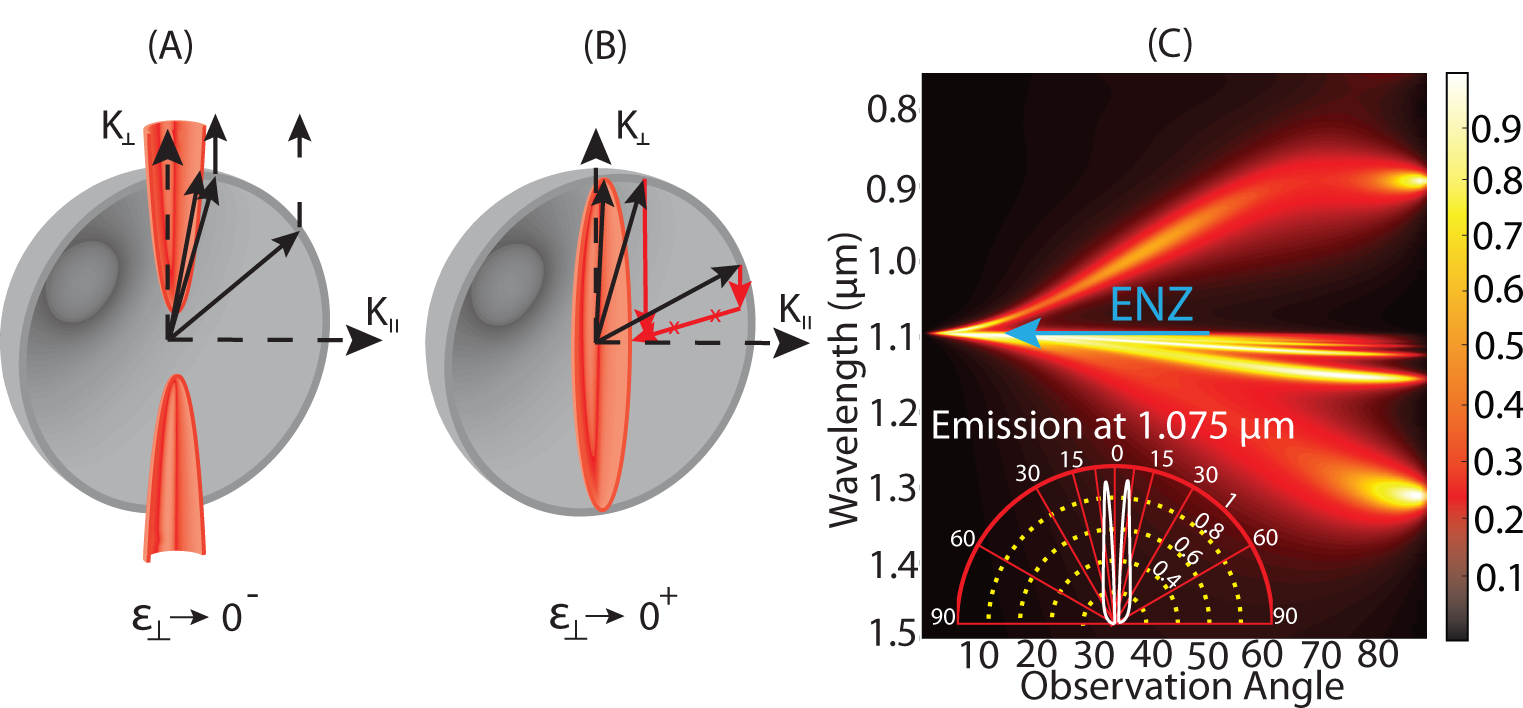}
\caption{Cross-sectional view of the narrow isofrequency surfaces of the metamaterial in the ENZ regime. The spherical isofrequency surface corresponds to vacuum. (A) As the perpendicular permittivity nears zero from the negative side, the dispersion relation inside the metamaterial becomes a narrow hyperboloid. (B) Ellipsoidal isofrequency surface as the perpendicular effective medium constant approaches ENZ from the positive side. Note that only waves at near-normal incidence from vacuum penetrate the metamaterial which are immediately absorbed due to the ENZ resonance. Furthermore, the large impedance mismatch at higher angles leads to high reflections. This results in highly directional emissivity patterns.  (C) P-polarized emissivity plot for a 450 nm thick metamaterial emitter consisting of a host matrix of aluminum oxide ($Al_{2}O_{3}$) embedded with 15 nm diameter silver nanowires in a 115 nm square unit cell using the effective medium approach. The angularly sharp emission near normal incidence around 1.075  $ \mu$m is usable for applications requiring coherent thermal radiation. The inset shows a polar plot of the emissivity along the 1.075  $ \mu$m line.  The secondary bands of high emissivity around the ENZ region is due to the impedance matching behavior in the ellispoidal/hyperboloidal isofrequency regime which moves to higher angles as the $|Re(\epsilon)\to 0|$ condition is relaxed. }
\end{figure}

\section{High temperature plasmonic metamaterials}
To this present time, optical metamaterial applications have been confined to temperatures near or significantly below 300 K. Yet, many of the most interesting applications of thermal engineering require temperatures in the vicinity of 1500 K. It follows directly that the scope of traditional optical metamaterials for thermal engineering applications is quite restricted. However, this limitation is strictly material based. In this section, we show that by switching away from the conventional low melting point metals, such as silver and gold, optical metamaterial design principles can be employed for high temperature thermal engineering.

\begin{minipage}{\linewidth}
\centering
\setlength{\tabcolsep}{20pt}
Tab. 1: Melting temperature and plasmonic figure of merit for near-IR metals.\par 
\bigskip
\begin{tabular}{c|c|c}{Material} &{Melting Point $\left(K\right)$} & {${-\epsilon^{'}/\epsilon^{''}} \left(1 \rightarrow 2 \mu m;  300 K\right)$} \\
\hline $Au$ & $1337\hspace{3mm}$ & $\hspace{4 mm}14 \rightarrow 8\hspace{4 mm}$ \cite{Johnson1972}\\
$Ag$ & $1235\hspace{3mm}$ & $\hspace{3 mm}98\rightarrow 29\hspace{3 mm}$ \cite{Johnson1972}\\ 
$AZO$ & $\approx 2200\hspace{1 mm} $\cite{Kim1997} & $\hspace{1.5 mm}-15\rightarrow 4 \hspace{3.5 mm}$\cite{Naik2010}\\
$TiN$ & $\hspace{3mm}3250\hspace{2 mm}$ \cite{Pierson1996} & $\hspace{5.2 mm}1\rightarrow 2 \hspace{5.2 mm}$\cite{Naik2011}
\label{MaterialTable}
\end{tabular}\par
\bigskip
\end{minipage}

Following the results of the above table, several thermally robust materials display weakly metallic optical characteristics and relatively low loss (Fig. 6(A)) through the 1 to 2  $ \mu$m spectral range. These characteristics prove crucial for thermal engineering. Building on the results of inset Fig.3(B), energy based applications of thermal engineering are most likely in the approximate temperature range of 1500 K. At significantly cooler temperatures, the spectrally broad and relatively small power density emitted by a blackbody (see Sec.V) strongly limits the energy conversion performance of any possible far field TPV device. Returning to the inset of Fig.3(B), since the bulk of the emitted power for a blackbody in this temperature range is positioned at wavelengths slightly longer than 1  $ \mu$m, this becomes the most important spectral region for emissivity control. 

According to the effective medium theory presented in previous sections, both ENP and ENZ resonances require the metallic and dielectric components of the unit cell to have $|Re\left(\epsilon\right)|$ of the same order. As $Re\left(\epsilon\right)$ is at most on the order of 10 for a dielectric in the near infrared spectral region, metamaterial based ENP and ENZ behavior in this region of high spectral power density is only possible if the metallic permittivity component satisfies $-10\leq Re\left(\epsilon\right)\leq -1$. This condition makes the replacement of silver or gold with more thermally robust interstitial nitrides such as titanium nitride \cite{Naik2011}, transition metals such as tantalum, or transparent conductive oxide semiconductors such as aluminum zinc oxide \cite{Naik2010}, all the more important for metamaterial thermal engineering. Fine-tuning of the ENP resonance using high temperature plasmonic materials is shown in Fig.6(B).
\begin{figure}
\centering
\includegraphics{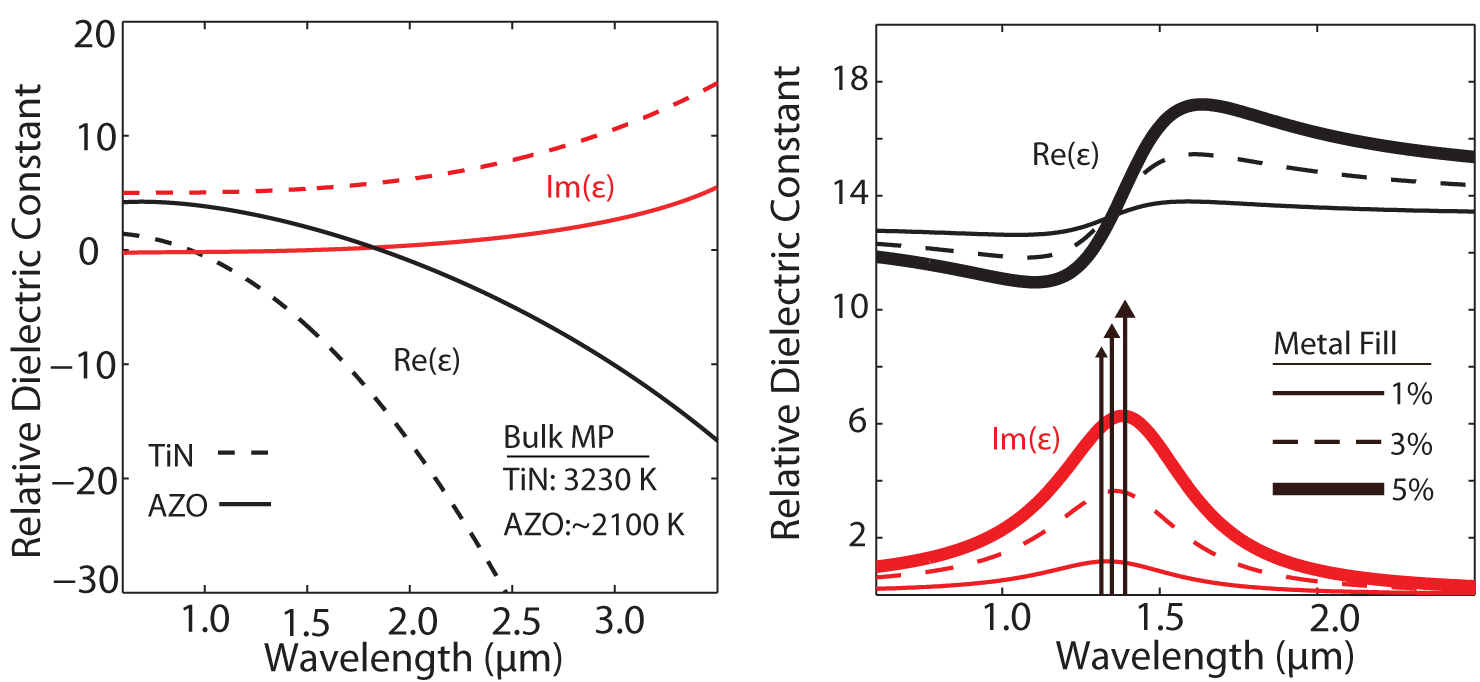}
\caption{ (A) Drude models of the optical properties of TiN and AZO based on the data presented in \cite{Naik2011,Naik2010}. (B) Fine-tuning of the ENP metamaterial resonance by altering the fill fraction of metal in the unit cell. In this plot the titanium nitride/silicon metamaterial system described in Fig.4 is used. Both AZO and TiN achieve thermally stable plasmonic behavior in the near infrared (Table Sec.4).}
\end{figure}
\section{Energy conversion efficiency of TPV devices}
\subsection{Characteristics of ENZ and ENP emitters for TPVs}
We now turn our attention to characterizing the performance of the metamaterial emitters made of high temperature plasmonic materials for TPV systems. We follow the theoretical energy conversion argument developed by Shockley and Queisser \cite{Shockley1961}. For completeness, we provide a detailed comparison with a blackbody emitter of the various efficiencies that need to be accounted for in a practical device.  

The model begins with the primary assumption that the photovoltaic cell, of the larger TPV device, operates with unity quantum efficiency perfectly cutoff at the bandgap energy. Under this assumption, photons with energy greater than the material bandgap create an electron hole carrier pair with energy equal to that of the bandgap, and any additional energy is lost to thermalization within the cell. Photons with energy below the bandgap of the photovoltaic cell are lost to free space. This perfect cutoff model defines an ultimate performance limitation on the TPV device following the usable power created by the emitter:
\begin{equation}
\eta_{ult}\left(\lambda_{gap},T\right)=\frac{\int_{0}^{\pi/2}cos\left(\theta\right)sin\left(\theta\right)d\theta \int_{0}^{\lambda_{g}}\frac{\lambda}{\lambda_{g}}\zeta_{E}\left(\lambda,\theta\right) I_{BB}\left(\lambda,T\right)d\lambda } {\int_{0}^{\pi/2} cos\left(\theta\right)sin\left(\theta\right)d\theta\int_{0}^{\infty}\zeta_{E}\left(\lambda,\theta\right) I_{BB}\left(\lambda,T\right)d\lambda}
\label{effult}
\end{equation}
\begin{equation}
I_{BB}\left(\lambda\right)=\frac{8\pi hc}{\lambda^{5} \left(e^{\frac{hc}{\lambda k_{B}T}}-1\right)},
\label{IBB}
\end{equation}
where $\zeta_{E}\left(\lambda,\theta\right)$ is emissivity of the emitter, $\lambda_{g}$ the wavelength of the material bandgap, $T$ the emitter temperature, and $I_{BB}$ the spectral radiance of an ideal blackbody. Here, have assumed the emitter to be planar with no azimuthal angular dependence and have used relative permittivities at 300 K. The high temperatures are not expected to alter the performance appreciably \cite{Zhang2000,Khurgin2012}.

To understand the consequences of this relation, the efficiencies $\eta_{ult}$ resulting from the angularly averaged emissivity profiles of an ideal blackbody and the titanium nitride/silicon nanowire system, described in Fig.4, are plotted for a gallium antimonide photovoltaic cell with temperatures varying from 500 to 2500 K in Fig.7(A). Gallium antimonide is chosen as the photovoltaic cell as its material bandgap, 0.71 eV, is well matched to the spectral power distribution of the blackbody near 1500 K.  At low temperatures, since the blackbody spectral power distribution is broad and flat, the power emitted near the photovoltaic cell bandgap is not an appreciable percentage of the total emitted power. As the temperature increases, the blackbody spectral radiance (Eq.(\ref{IBB})) increases and shifts towards shorter wavelengths. This leads to higher percentages of the emitted radiation being usable by the photovoltaic cell and, consequently, increased ultimate efficiency. This effect peaks as the efficiency gained by shifting a larger portion of the spectral power density to lower wavelengths is countered by mounting thermal losses due to the cell only producing electron hole pairs at the bandgap energy. A visualization of the spectral power distributions of an ideal black body, the titanium nitride nanowire design, and the emitter offering peak device efficiency for a gallium antimonide photovoltaic cell are shown in Fig.7(B) at a temperature of 1500 K. 
\begin{figure}
\centering
\includegraphics{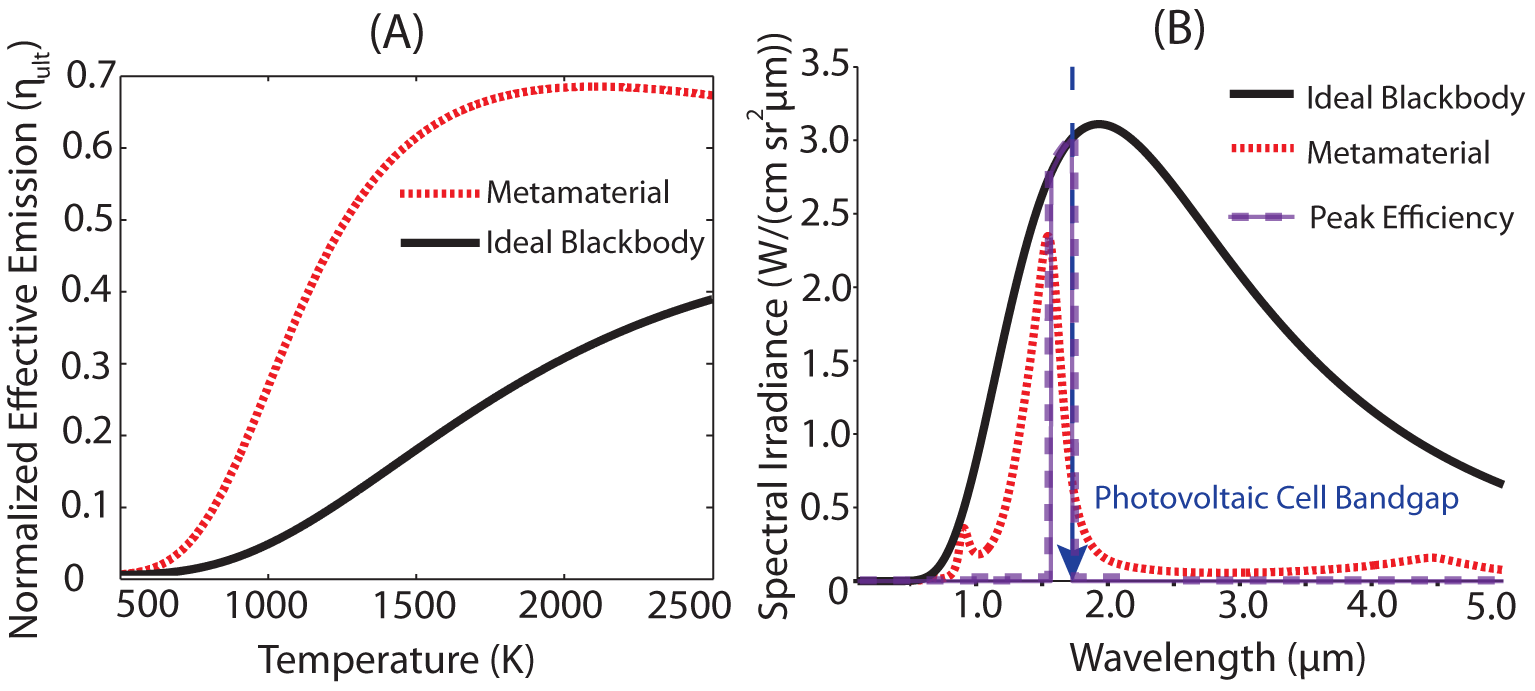}
\caption{ (A) Comparison of the ultimate efficiency of a titanium nitride metamaterial emitter (Fig.4) to that of a blackbody for a 0.71 eV material bandgap, corresponding to GaSb. Based on bulk material parameters, the metamaterial emitter will be thermally stable up to 1650 K. (B) Comparison of the angularly averaged spectral emission characteristics between the titanium nitride metamaterial design, an ideal blackbody, and an emitter which maximizes the efficiency of energy conversion at 1500 K. The large lobe of the metamaterial ENP resonance closely matches the position and magnitude of the emitter producing the highest TPV device efficiency.}
\end{figure}
\subsection{Detailed balance efficiency}

To determine the total efficiency of our emitter designs as part of a TPV device, we take into account the extra efficiency terms corresponding to 1) the additional ways in which charge carriers may be removed from the cell and 2) specific current voltage relations of a standard pn-junction.

In addition to the desired external current, both radiative and nonradiative recombinations reduce the total number of charge carriers within a photovoltaic cell. The characteristics of these recombination mechanisms can be understood through the equilibrium condition of the photovoltaic cell with and without the presence of emitter photons. With no emitter contribution, and the cell in thermodynamic equilibrium, the net number of carriers within the cell is constant, and can be described by:
\begin{equation}
Q_{BB}\left(\lambda_{g},T_{C}\right)-R_{rad}\left(T_{C}\right)+G_{other}\left(T_{C}\right)-R_{other}\left(T_{C}\right)=0,
\end{equation}
where $Q_{BB}\left(T_{C}\right)$ is the number of carriers generated by photons incident on the photocell from its surroundings,
\begin{equation}
Q_{BB}\left(\lambda_{g},T_{C}\right)=2 \int_{0}^{2\pi}d\phi\int_{0}^{\pi/2}cos\left(\theta\right)sin\left(\theta\right)d\theta\int_{0}^{\lambda_{g}}\frac{\lambda}{\lambda_{g}} I_{BB}\left(\lambda,T_{C}\right)d\lambda,
\label{QBB}
\end{equation}
$R_{rad}\left(T_{C}\right)$ the number lost by radiative recombination, $G_{other}\left(T_{C}\right)$ any other carrier generation, and $R_{other}\left(T_{C}\right)$ any other carrier recombination. As this equilibrium also requires that energy entering the cell is equal to that exiting, we can immediately conclude that:
\begin{equation}
R_{rad}\left(T_{C}\right)=Q_{BB}\left(\lambda_{g},T_{C}\right)\hspace{10 mm}R_{other}\left(T_{C}\right)=G_{other}\left(T_{C}\right).
\end{equation}
We now examine the cell with an external incident flux from an emitter source, and an external load through which carriers may exit. Following the standard statistical mechanics model, we assume that the additional carrier perturb the recombination rates within the cell as $e^{V/V_{C}}$. Here, V is the difference between the quasi-Fermi levels, and $V_{C}=\frac{k_{B}T_{C}}{e}$. As such,
\begin{equation}
Q_{E}\left(\lambda_{g},T_{E}\right)-Q_{BB}\left(\lambda_{g},T_{C}\right) e^{\frac{V}{V_{C}}}+G_{other}\left(T_{C}\right)-G_{other}\left(T_{C}\right)e^{\frac{V}{V_{c}}}-\frac{I}{q}=0,
\label{balance}
\end{equation}
\begin{equation}
Q_{E}\left(\lambda_{g},T_{E}\right)=\int_{0}^{2\pi}d\phi\int_{0}^{\pi/2}cos\left(\theta\right)sin\left(\theta\right)d\theta \int_{0}^{\lambda_{g}}\frac{\lambda}{\lambda_{g}}\zeta_{E}\left(\lambda,\theta\right) I_{BB}\left(\lambda,T_{E}\right)d\lambda,
\label{QE}
\end{equation}
as we assume the emitter photons fall on only one side of the photovoltaic cell. Note that the generation term is primarily due to the high temperature emitter (T=1500K) and the finite volume between the emitter and photovoltaic cell has a minimal contribution to the incident radiation. Furthermore, following standard TPV efficiency calculations, we assume that any electromagnetic energy radiated by the solar cell is not reused for energy conversion. Introducing $f_{rec}$ as the percentage of recombinations which are radiative $\left(\frac{Q_{BB}\left(\lambda_{g},T_{C}\right)}{Q_{BB}\left(\lambda_{g},T_{C}\right)+G_{other}\left(T_{C}\right)}\right)$ and solving for the open circuit voltage we find:
\begin{equation}
V_{OC}=V_{C}\hspace{1 mm}ln\left(\frac{f_{rec}}{2}\frac{Q_{E}\left(\lambda_{g},T_{E}\right)}{Q_{BB}\left(\lambda_{g},T_{C}\right)}-f_{rec}+1\right),
\end{equation}
which, when functioning within the quasi-static limit of this approximation, is smaller than the initial material bandgap $V_{g}$. This consideration leads us to the efficiency factor:
\begin{equation}
\eta_{rec}\left(\lambda_{g},T_{E},T_{C}\right)=\frac{V_{OC}}{V_{g}}.
\label{effrec}
\end{equation}
Returning to Eq. (\ref{balance}), we can see that a final efficiency reduction arises due to the interrelation of current and voltage in the pn-junction cell. As current is allowed to flow the number of free carriers within the cell drops, reducing the voltage from its open circuit value. As current must be drawn to create usable power, an efficiency factor relating the operational voltage ($V_{P}$) and current ($I\left(V_{P}\right)$) to the open circuit voltage ($V_{OC}$) and short circuit current ($I_{SC}$) must be included by finding the maximal power-point ($V_{P}I(V_{P})$) of the photovoltaic cell. Using (\ref{balance}) we find:
\begin{equation}
\eta_{pow}=\frac{V_{P}I\left(V_{P}\right)}{V_{OC}I_{SC}}=\frac{v^{2}_{pp}}{\left(v_{pp}+ln\left(1+v_{pp}\right)\right)\left(1+v_{pp}-e^{-v_{pp}}\right)},
\label{effpow}
\end{equation}
where $v_{PP}$ is defined as the ratio of the operating cell voltage at the maximum power-point to the background cell voltage, $\frac{V_{P}}{V_{C}}$, and is determined via the relation:
\begin{equation}
V_{OC}=V_{P}+V_{C}\hspace{1 mm}ln\left(1+\frac{V_{P}}{V_{C}}\right).
\end{equation}
The product of these two results, $\eta_{rec}\eta_{pow}$, shows the efficiency with which an idealized pn-junction photovoltaic cell converts photons at its material bandgap into usable power. Using the three efficiency parameters computed, we can now determine the overall efficiency of our metamaterial emitter designs functioning as part of a TPV system.
\subsection{Energy conversion results}
\begin{figure}[ht!]
\centering
\includegraphics{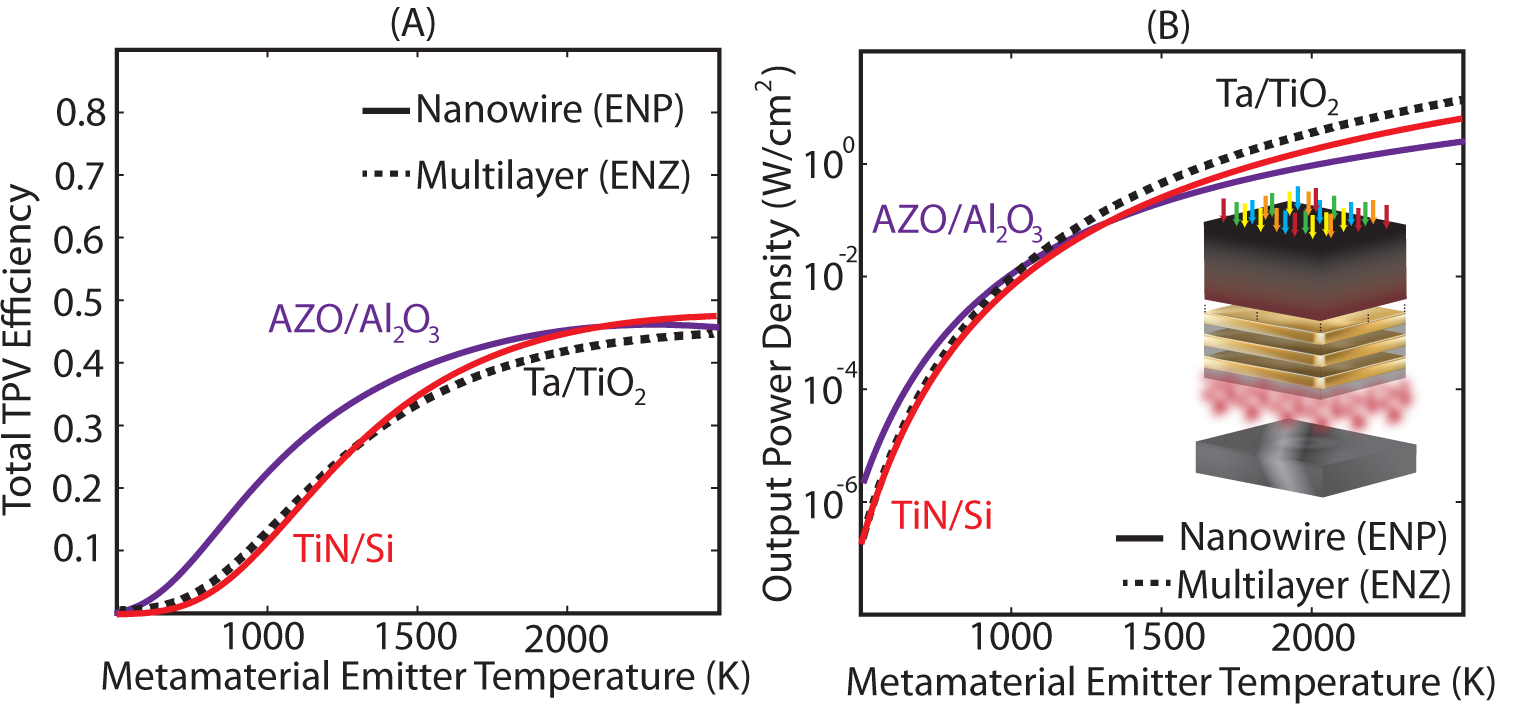}
\caption{ (A) Theoretical efficiency of three TPV devices taking into account all discussed effects. The cell parameters of the $Ta/TiO_{2}$ multilayer and $TiN/Si$ nanowire systems are the same as in Fig. 3 and Fig. 4. The third system utilizes 250 nm long, 20 nm diameter AZO rods in a 125 nm square $Al_{2}O_{3}$ matrix, set on an optically thick tantalum backing, and an InGaAs photovoltaic cell with bandgap set at 2100 nm. In these plots the efficiency of heating the source is not included. However, due to the tantalum backing included in all designs, the performance of these devices should not be greatly altered by the characteristics of the heat source. (B) Final output power density showing the potential for TPV. Due to the lower energy bandgap of the InGaAs photovoltaic cell, the AZO based metamaterial system produces relatively higher power density at lower temperatures. The opposite is seen at higher temperatures. (inset) Schematic of a multilayer metamaterial near the TPV cell.}
\end{figure}
From the ultimate efficiency (Eq.(\ref{effult}) and Fig.7) we can begin to see that the principal characteristic of a good emitter for TPVs is the suppression of subbandgap photons. Near unity ultimate efficiency, $\eta_{ult}$, requires the spectral emissivity of the emitter, $\zeta_{E}\left(\lambda,\theta\right)$, to become needle-like at a wavelength slightly shorter than that of the material bandgap. Yet, this ultra-sharp spectral behavior is pragmatically undesirable. As the spectral range of emission is narrowed, the power produced at a fixed temperature is decreased. This reduction not only limits the amount of power that can be extracted from the TPV device, but also acts to reduce the total energy conversion efficiency. Reviewing  Eq.(\ref{effrec}), if the number of carriers in the pn-junction is near the unilluminated thermal equilibrium amount, a large percentage of the incident emitter energy will be either reradiated or lost within the cell. Consequently, the efficiency of converting the power radiating from the emitter will also be reduced. Since the temperature of the emitter, and thus the amount of usable power it can produce, is intrinsically limited by the materials used, whether a spectrally wider or narrower emission is preferable depends on the particular application. At low temperatures, broader emitters are more efficient and provide greater power densities; however, as temperature is increased their efficiency falls below that of the narrow emitter.

Clearly, this tradeoff has direct consequence for our TPV emitter designs. For applications in which it is not possible to reach emitter temperatures of $\approx$ 1000 K, broad ENZ type emitters will create better TPV devices. Above this approximate temperature, spectrally thinner ENP type absorption peaks will provide superior net efficiency. Here, even lower emitted power begins to become a benefit of the spectrally thin design. As the emitter is in thermodynamic equilibrium, all emitted energy must be replaced. This becomes increasingly difficult with a spectrally broad emitter at high temperatures. As an alternative view, given a fixed input power, a spectrally thinner emitter will operate at a higher temperature and thus higher efficiency (Fig.8(B)).

Taking all three efficiency considerations into account, the theoretical metrics for TPV devices using metamaterial emitters, neglecting any loss associated with an absorber stage, are shown in Fig.8. The full concentration limit for solar based single junction photovoltaics, following the assumptions of Shockley and Queisser, is surpassed for emitter temperatures near 1500 K in the AZO nanowire system. The overall performance of these designs show either greater high temperature efficiency \cite{Wu2012,Celanovic2004}, tighter spectral emissivity, or similar efficiency behavior with a higher emitter power \cite{Rephaeli2009} to other contemporary designs.

\section{Concluding remarks}
In this article we have introduced a class of artificial media called epsilon-near-pole metamaterials and developed a general means of exerting control over thermally excited far field electromagnetic radiation through the use of ENP and ENZ resonances. Thermal radiation control using these generalized bulk material resonances shows great potential for creating a new class of realizable thermal devices. In particular, we have shown specific metamaterial implementations for creating emitters for TPVs and coherent thermal sources in the near infrared range crucial for energy applications. We have also introduced the concept of metamaterials based on high temperature plasmonic materials. This switch away from typical plasmonic materials is crucial for many far field thermal applications and more generally for furthering the scope of optical metamaterial designs. This work paves the way for future use of metamaterials for the control of thermally excited radiation. 
\section*{Acknowledgments}
We would like to acknowledge funding from the Canadian National Science and Engineering Research Council (NSERC), Canadian School of Energy and Environment (CSEE) and Alberta Nanobridge.


\end{document}